\documentclass{article}
\usepackage{amssymb}

\usepackage{graphicx}
\usepackage{amsmath}

\newtheorem{theorem}{Theorem}
\newtheorem{acknowledgement}[theorem]{Acknowledgement}

\input{tcilatex}

\begin{document}

\title{From On-shell to Off-shell Open Gauge Theories\thanks{%
Final version, accepted for publication in Physical Review D.}}
\author{N. Djeghloul\thanks{%
E-mail: ndjeghloul@univ-oran.dz} \ and M. Tahiri \\
Laboratoire de Physique Th\'{e}orique\\
Universit\'{e} d'Oran Es-Senia, 31100 Oran, Algeria}
\date{\today}
\maketitle

\begin{abstract}
We present an alternative quantization for irreducible open gauge theories.
The method relies on the possibility of modifying the classical BRST
operator and the gauge-fixing action written as in Yang-Mills type theories,
in order to obtain an on-shell invariant quantum action by using equations
characterizing the full gauge algebra. From this follows then the
construction of an off-shell version of the theory. We show how it is
possible to build off-shell BRST algebra together with an invariant
extension of the classical action. This is realized via a systematic
prescription for the introduction of auxiliary fields.

\end{abstract}

\newpage

\section{Introduction}

It is well known that an on-shell quantization of general gauge theories,
i.e. gauge theories which are reducible and/or whose classical gauge algebra
is closed only on-shell (for a review see Ref. [1]), can successfully
realized in the Lagrangian approach by the Batalin-Vilkovisky (BV) formalism 
\cite{2}.

In this framework, the field content of the theory is doubled by the
introduction of the so-called anti-fields. The procedure consists, through
the elimination of the antifields via a gauge-fixing fermion of ghost number 
$(-1)$, in the construction of the quantum theory in which the effective
BRST transformations are nilpotent on-shell.

Let us note that the BV approach is not the only alternative to quantize
reducible and/or open gauge theories. Indeed, the introduction of a set of
auxiliary fields, as in supersymmetric theories \cite{3} or in BF theories 
\cite{4}, may close the gauge algebra, and then gives the possibility to use
the standard BRST formalism in the context of the Faddeev-Popov procedure 
\cite{5}.

However, no systematic prescription exists in order to introduce these
auxiliary fields so that an approach that will be able to realize the
on-shell as well as the off-shell quantization of general gauge theories in
a systematic way will appear to be superior to all other available schemes.

Recently \cite{6} we show for the case of simple supergravity how an
on-shell quantization approach of the theory can lead, via a convenient
procedure, to find out the structure of auxiliary fields as well as the full
BRST operator that realize off-shell quantization of the theory. The aim of
the present paper is to extend the analysis developed in Ref. [6] in order
to discuss general irreducible open gauge theories, irrespective of the
underlying classical action.

The paper is organized as follows: In Sec. \ref{II} we perform on-shell
quantization for a general irreducible open gauge theory by using the
structure of the gauge algebra. This is a new more natural quantization
procedure, in the sense that we will not relying on any set of extra fields.
Sec. \ref{III} is divided into two subsections. In the fist one we show how
it is possible to introduce a set of auxiliary fields to build the full
off-shell quantum action and the associated off-shell BRST symmetry for the
case of irreducible open gauge theories of type (2,2). The invariant
extension of the classical action is also given. In the next subsection, a
complete generalization is given. In Sec. \ref{IV} the specific problem of
the construction of the minimal set of auxiliary fields for any given
irreducible theory is analyzed. Section \ref{V} is devoted to concluding
remarks.

\section{\label{II}On-shell Quantization}

Let us consider an arbitrary gauge theory whose classical action $S(\Phi
^{i})$ possesses local gauge symmetries 
\begin{equation}
\Delta S=0,  \tag{1}
\end{equation}
with 
\begin{equation}
\Delta \Phi ^{i}=(-)^{i\alpha }R_{\alpha }^{i}\varepsilon ^{\alpha }, 
\tag{2}
\end{equation}
where $\{\Phi ^{i},i=1,...,N\}$ describes the set of classical fields of the
theory and the operators $R_{\alpha }^{i}$ are acting on the parameters $%
\{\varepsilon ^{\alpha },\alpha =1,...,d\}$ of the $d$ symmetries of $S$ and 
$i(\alpha )$ is the parity of $\Phi ^{i}(\varepsilon ^{\alpha }).$ The
invariance condition (1) leads to the Noether's identity 
\begin{equation}
R_{\alpha }^{i}\frac{\delta S}{\delta \Phi ^{i}}=0.  \tag{3}
\end{equation}
Dealing with irreducible symmetries \cite{1}, we also have 
\begin{equation}
\forall X_{A}^{\alpha }:\text{ }R_{\alpha }^{i}X_{A}^{\alpha }=0\Rightarrow
X_{A}^{\alpha }=0,  \tag{4}
\end{equation}
where $A$ represents an arbitrary set of indices.

The condition (3) allows to define $d$ operators $\Delta _{\alpha }$%
\begin{equation}
\Delta _{\alpha }\Phi ^{i}=R_{\alpha }^{i},  \tag{5}
\end{equation}
which satisfy 
\begin{equation}
\Delta _{\alpha }S=0.  \tag{6}
\end{equation}
The graded commutator of two transformations is then given by 
\begin{equation}
\left[ \Delta _{\alpha },\Delta _{\beta }\right] \Phi ^{i}=R_{\alpha }^{j}%
\frac{\delta R_{\beta }^{i}}{\delta \Phi ^{j}}-(-)^{\alpha \beta }R_{\beta
}^{j}\frac{\delta R_{\alpha }^{i}}{\delta \Phi ^{j}}.  \tag{7}
\end{equation}
Considering that the set of the $R_{\alpha }^{i}$ is complete, i.e. all the
symmetries of $S$ are known, one can easily find that the most general form
of the gauge algebra reads \cite{1} 
\begin{equation}
\left[ \Delta _{\alpha },\Delta _{\beta }\right] \Phi ^{i}=T_{\alpha \beta
}^{\lambda }R_{\lambda }^{i}+V_{\alpha \beta }^{ij}\frac{\delta S}{\delta
\Phi ^{j}}.  \tag{8}
\end{equation}
Therefore, the properties of the gauge algebra will depend on the nature of
the structure functions $T_{\alpha \beta }^{\lambda }$ and the non closure
functions $V_{\alpha \beta }^{ij}$, which depend in general on the classical
fields and are graded antisymmetric with respect to $(\alpha \beta )$ and $%
(ij).$

In view of Eq. (8), the generalized graded Jacobi identity can be put in the
form 
\begin{eqnarray}
&&\sum_{(\alpha \beta \gamma )}\{R_{\alpha }^{k}T_{\beta \gamma ,k}^{\lambda
}R_{\lambda }^{i}-(-)^{\alpha (\beta +\gamma )}T_{\beta \gamma }^{\sigma
}T_{\sigma \alpha }^{\lambda }R_{\lambda }^{i}+\{R_{\alpha }^{k}V_{\beta
\gamma ,k}^{ij}-  \TCItag{9} \\
&&(-)^{\alpha (\beta +\gamma )}((-)^{\alpha i}V_{\beta \gamma
}^{ik}R_{\alpha ,k}^{j}+(-)^{ij+1}(-)^{\alpha j}V_{\beta \gamma
}^{jk}R_{\alpha ,k}^{i}+T_{\beta \gamma }^{\sigma }V_{\sigma \alpha
}^{ij})\}S_{,i}\}=0,  \notag
\end{eqnarray}
where $\sum_{(\alpha \beta \gamma )}$ means a cyclic sum over $\alpha ,$ $%
\beta ,$ $\gamma $ and $``,k"$ means a variation with respect to $\Phi ^{k}.$

However, the standard BRST approach consists in the replacement of the local
gauge invariance by a global one. This symmetry is encoded in an operator $%
\delta $ defined via the replacement of the gauge parameters $\varepsilon
^{\alpha }$ by the ghost fields $c^{\alpha }$ with parity $(\alpha +1)$ and
ghost number $(+1)$, we have 
\begin{equation}
\delta \Phi ^{i}=(-1)^{i(\alpha +1)}R_{\alpha }^{i}c^{\alpha },  \tag{10}
\end{equation}
which maintains the classical action invariant.

It is easy to show that the action of $\delta $ on $\Phi ^{i}$ is nilpotent
on-shell, so that 
\begin{equation}
\delta ^{2}\Phi ^{i}=V^{ij}S_{,j},  \tag{11}
\end{equation}
where 
\begin{equation*}
V^{ij}=\frac{1}{2}(-)^{\beta (\alpha +1)}(-)^{(i+j)(\alpha +\beta
)}V_{\alpha \beta }^{ij}c^{\alpha }c^{\beta }\text{ ,}
\end{equation*}
provided that the transformation of the ghost is given by 
\begin{equation}
\delta c^{\lambda }=-\frac{1}{2}(-)^{\beta (\alpha +1)}(-)^{\lambda (\alpha
+\beta )}T_{\alpha \beta }^{\lambda }c^{\alpha }c^{\beta },  \tag{12}
\end{equation}
which is also nilpotent on-shell. Indeed, by using the graded Jacobi
identity, we obtain 
\begin{equation}
R_{\lambda }^{i}\delta ^{2}c^{\lambda }=(-)^{i(\lambda +1)}\{\delta
V^{ij}-((-)^{i+j(\lambda +1)}V^{ik}R_{\lambda ,k}^{j}c^{\lambda
}+(-)^{ij+1}(i\rightleftharpoons j))\}S_{,j}.  \tag{13}
\end{equation}
This means that $R_{\lambda }^{i}\delta ^{2}c^{\lambda }$ vanishes on-shell
and because $R_{\lambda }^{i}$ describes irreducible transformations, then $%
\delta ^{2}c^{\lambda }$ also vanishes on-shell and can be cast in the form 
\begin{equation}
\delta ^{2}c^{\alpha }=Z^{\alpha j}S_{,j},  \tag{14}
\end{equation}
where the new non closure functions $Z^{\alpha j}$ satisfy Eq. (13). This
characteristic equation represents the fact that $Z^{\alpha j}$ are not
completely independent from $V^{ij}$. It can also be derived by acting $%
\delta $ on Eq. (11) and written as 
\begin{eqnarray}
&&\{\delta V^{ij}-((-)^{j(i+k+1)}V^{kj}\left( \delta \Phi ^{i}\right) _{,k}+
\notag \\
&&(-)^{j(\alpha +1)}Z^{\alpha j}\left( \delta \Phi ^{i}\right) _{,\alpha
}+(-)^{ij+1}(i\rightleftharpoons j))\}S_{,j}=0,  \TCItag{15}
\end{eqnarray}
where $``,\alpha "$ means a variation with respect to $c^{\alpha }$. One can
remark that the above equation is of the third order in ghost, and indicates
the possibility of existence of a new characteristic function $V^{ijk}$
defined by 
\begin{eqnarray}
&&\delta V^{ij}-((-)^{j(i+k+1)}V^{kj}\left( \delta \Phi ^{i}\right) _{,k}+ 
\notag \\
&&(-)^{j(\alpha +1)}Z^{\alpha j}\left( \delta \Phi ^{i}\right) _{,\alpha
}+(-)^{ij+1}(i\rightleftharpoons j))=V^{ijk}S_{,k},  \TCItag{16}
\end{eqnarray}
where $V^{ijk}$ are restricted by the total graded antisymmetry, $%
V^{ijk}=(-)^{ij+1}V^{jik}=(-)^{kj+1}V^{ikj}.$

We can also introduce a function $Z^{\alpha ij}$ from the ghost non closure
function $Z^{\alpha i}$ by acting $\delta $ on Eq. (14) and find then the
following characteristic equation 
\begin{eqnarray}
&&\delta Z^{\alpha i}-(-)^{i(\alpha +\beta +1)}Z^{\beta i}(\delta c^{\alpha
})_{,\beta }-  \notag \\
&&(-)^{i(\alpha +k)}V^{ki}(\delta c^{\alpha })_{,k}+(-)^{\alpha +1}Z^{\alpha
k}\left( \delta \Phi ^{i}\right) _{,k}=Z^{\alpha ij}S_{,j},  \TCItag{17}
\end{eqnarray}
where $Z^{\alpha ij}=(-)^{ij+1}Z^{\alpha ji}$.

It is worth noting that an other application of $\delta $ on Eq. (16) (Eq.
(17)) leads to an equation which allows to introduce an other function of
type $V^{ijkl}$ $\left( Z^{\alpha ji}\right) $, and so on for all orders of
application of $\delta $. The general characteristic functions produced in
this way are all related by equations derived in the same way as Eqs. (16)
and (17). We denote the characteristic functions defined from an equation of
order $n$ in application of $\delta $ by $V_{n}^{i_{1}...i_{n}}$ and $%
Z_{n}^{\alpha i_{1}...i_{n-1}}$. They are graded antisymmetric with respect
to the indices $i_{l}$ $(l=1,...,n-1,n)$. At an order $(n+1)$ we find the
following characteristic equations 
\begin{eqnarray}
&&\delta
V_{n}^{i_{1}...i_{n}}-\sum_{m=2}^{n}(-)^{m}%
\{V_{m}^{ki_{n-m+2}...i_{n}}(V_{n-m+1}^{i_{1}...i_{n-m+1}})_{,k}- 
\TCItag{18} \\
&&Z_{m}^{\alpha i_{n-m+2}...i_{n}}(V_{n-m+1}^{i_{1}...i_{n-m+1}})_{,\alpha
}\}=V_{n+1}^{i_{1}...i_{n+1}}S_{,i_{n+1}},  \notag
\end{eqnarray}
\begin{eqnarray}
&&\delta Z_{n}^{\alpha i_{1}...i_{n-1}}-\sum_{m=2}^{n}(-)^{m}\{Z_{m}^{\beta
i_{n-m+1}...i_{n-1}}(Z_{n-m+1}^{\alpha i_{1}...i_{n-m}})_{,\beta }- 
\TCItag{19} \\
\text{ } &&Z_{m}^{\alpha
i_{1}...i_{m-2}k}(V_{n-m+1}^{i_{m-1}...i_{n-1}})_{,k}+V_{m}^{ki_{n-m+1}...i_{n-1}}(Z_{n-m+1}^{\alpha i_{1}...i_{n-m}})_{,k}\}=Z_{n+1}^{\alpha i_{1}...i_{n}}S_{,i_{n}},
\notag
\end{eqnarray}
where graded antisymmetrization over all independent combinations related to
the indices $(i_{1},...,i_{n})$ must be carried out. Note that the functions 
$V_{n}$ and $Z_{n}$ have parity $(i_{1}+...+i_{n}+n$ $mod2)$ and $(\alpha
+i_{1}+...+i_{n-1}+n$ $mod2)$ and ghost numbers $\left( n\right) $ and $%
\left( n+1\right) $, respectively.

The existence of these characteristic functions $V_{n}$ and $Z_{n}$ permits
a classification for irreducible open gauge theories. We will say that a
theory is of type $(p,q)$ in the case where $V_{n}=0$ $(Z_{n}=0)$ for $n>p$ $%
(n>q)$. For example, global supersymmetric theories as well as
Super-Yang-Mills theories are of type $(2,1)$ while simple supergravity is
of type $(2,2)$.

In what follows we turn to discuss how to construct the quantum theory of a
classical open gauge theory of type $(p,q)$. It is obvious that a $\delta $%
-exact form of the gauge fixing action cannot be suitable to build the full
invariant quantum action, because of the on-shell nilpotency of the BRST
operator $\delta $. To this end, we generalize the prescription discussed in
Ref. 6 for the case of simple supergravity by simply modifying the classical
BRST operator $\delta $. As a consequence the gauge-fixing action written as
in Yang-Mills theories must be also modified, so that the complete quantum
action becomes invariant. We first introduce the gauge fermion $\Psi $ of
ghost number $(-1)$ to implement the gauge constraints $F_{\alpha }=0$
associated to all the invariances of the classical action $S$, we have 
\begin{equation}
\Psi =\bar{c}^{\alpha }F_{\alpha },  \tag{20}
\end{equation}
where $\bar{c}^{\alpha }$ $(\alpha =1,...,d)$ represent the antighosts with
parity $(\alpha +1)$ and ghost number $(-1)$, which allow as usual to define
the Stueckelberg auxiliary fields $b^{\alpha }$ through the action of the
transformation $\delta $, so that 
\begin{equation}
\delta \bar{c}^{\alpha }=b^{\alpha },\text{ \quad }\delta b^{\alpha }=0. 
\tag{21}
\end{equation}
Let us note that the gauge-fixing functions $F_{\alpha }$ depend only on the
classical fields $\Phi ^{i}$, since the gauge symmetries are considered as
irreducible.

At the quantum level we have to define a modified BRST operator $Q$. This
will be done by introducing a set of operators $\delta _{n}$ given by 
\begin{eqnarray}
\delta _{0}\Phi ^{i} &=&\delta \Phi ^{i},  \TCItag{22-a} \\
\delta _{n}\Phi ^{i} &=&\frac{1}{n!}(-)^{in+a_{n}}V_{n+1}^{ii_{1}...i_{n}}%
\Psi _{,i_{1}}...\Psi _{,i_{n}}\text{\quad }n=1,...,p-1,  \TCItag{22-b}
\end{eqnarray}
for the classical fields, and 
\begin{eqnarray}
\delta _{0}c^{\alpha } &=&\delta c^{\alpha },  \TCItag{23-a} \\
\delta _{n}c^{\alpha } &=&\frac{1}{n!}(-)^{(\alpha
+1)n+a_{n}}Z_{n+1}^{\alpha i_{1}...i_{n}}\Psi _{,i_{1}}...\Psi _{,i_{n}}%
\text{ \quad }n=1,...,q-1,  \TCItag{23-b}
\end{eqnarray}
where $a_{n}=\sum_{r=2}^{n}i_{r}\sum_{s=1}^{r-1}(i_{s}+1)$ gives to $%
(-)^{a_{n}}\Psi _{,i_{1}}...\Psi _{,i_{n}}$ the same graded symmetry
properties than $V_{n+1}^{ii_{1}...i_{n}}$ and $Z_{n+1}^{\alpha
i_{1}...i_{n}}$. For the other fields $\bar{c}^{\alpha }$ and $b^{\alpha }$
the action of the $\delta _{n}$ operators is taken to be trivial, i.e. $%
\delta _{0}\bar{c}^{\alpha }=\delta \bar{c}^{\alpha }$, $\delta
_{0}b^{\alpha }=\delta b^{\alpha }$ and $\delta _{n}\bar{c}^{\alpha }=\delta
_{n}b^{\alpha }=0$ for $n>0$. We are now able to define the effective BRST
operator $Q$ 
\begin{eqnarray}
Q\Phi ^{i} &=&\sum_{n=0}^{p-1}\delta _{n}\Phi ^{i},\text{ \quad }Qc^{\alpha
}=\sum_{n=0}^{q-1}\delta _{n}c^{\alpha },  \TCItag{24-a} \\
Q\bar{c}^{\alpha } &=&\delta _{0}\bar{c}^{\alpha },\text{ \qquad \ }%
Qb^{\alpha }=\delta _{0}b^{\alpha },  \TCItag{24-b}
\end{eqnarray}
which leaves invariant the following full quantum action $S_{q}$%
\begin{equation}
S_{q}=S+\sum_{n=0}^{p-1}\frac{1}{n+1}\delta _{n}\Psi .  \tag{25}
\end{equation}
The first term $(n=0)$ of the gauge-fixing action, $S_{gf}=\sum_{n=0}^{p-1}%
\frac{1}{n+1}\delta _{n}\Psi $, leads to the standard result of the
Yang-Mills type theories while the other terms describe higher ghost
couplings which characterize open gauge theories. To prove the invariance of
the quantum action (25) under the effective BRST symmetry defined by
(24-a,b) we take advantage of the characteristic equations (18) and (19)
together with the on-shell nilpotency (11) and (14) of the classical BRST
operator $\delta $.

Furthermore, using again the characteristic equations (18) and (19), we find
that the effective BRST operator $Q$ is nilpotent on-shell at the quantum
level, i.e. with respect to the quantum equations of motion derived from the
quantum action (25). Indeed, we have 
\begin{eqnarray}
Q^{2}\Phi ^{i} &=&A^{ik}S_{q,k}+B^{\alpha i}S_{q,\alpha },  \TCItag{26-a} \\
Q^{2}c^{\alpha } &=&B^{\prime \alpha i}S_{q,i},  \TCItag{26-b} \\
Q^{2}\bar{c}^{\alpha } &=&Q^{2}b^{\alpha }=0,  \TCItag{26-c}
\end{eqnarray}
where 
\begin{eqnarray}
A^{ik} &=&\sum_{n=1}^{p-1}\frac{(-)^{n-1}}{(n-1)!}%
(-)^{(i+k)(n+1)+a_{n-1}}V_{n+1}^{iki_{1}...i_{n-1}}\Psi _{,i_{1}}...\Psi
_{,i_{n-1}},  \TCItag{27-a} \\
B^{\alpha i} &=&-\sum_{n=1}^{q-1}\frac{(-)^{\alpha (n+1)}}{(n-1)!}%
(-)^{i(n+\alpha )+a_{n-1}}Z_{n+1}^{\alpha ii_{1}...i_{n-1}}\Psi
_{,i_{1}}...\Psi _{,i_{n-1}},  \TCItag{27-b} \\
B^{\prime \alpha i} &=&\sum_{n=1}^{q-1}\frac{1}{(n-1)!}(-)^{(\alpha
+i)(n+1)+a_{n-1}}Z_{n+1}^{\alpha ii_{1}...i_{n-1}}\Psi _{,i_{1}}...\Psi
_{,i_{n-1}}.  \TCItag{27-c}
\end{eqnarray}
It is remarkable that the used prescription, which simply consists in the
modification of the classical BRST operator and of the gauge-fixing action
written as in Yang-Mills theories, provides a natural on-shell quantization
scheme for open irreducible gauge theories in the sense that it does not
need to rely on any set of extra fields (such as antifields).

\section{\label{III}Off-shell Quantization}

We are going now to discuss how we can introduce auxiliary fields, as
generalization of the approach developed in Ref. [6], so that we end up with
an off-shell structure for open gauge theories. To this end, and for the
sake of the procedure, we perform first the generalization for classical
open gauge theories of type $(2,2)$, then a complete generalization will be
straightforwardly given.

\subsection{Open gauge theories of type $(2,2)$}

In this case the theory is only characterized by the functions $V^{ij}$ and $%
Z^{\alpha i}$ and all the remaining characteristic functions $V_{n}$ and $%
Z_{n}$ for $n>2$ vanish. Also for simplicity and to present computations
leading to insight in the generalization of the analysis in Ref. 6 to open
gauge theories, we consider an open gauge algebra of type $(2,2)$ in which
the classical degrees of freedom $(\Phi ^{i})$ as well as the different
parameters of the classical symmetry $\left( \varepsilon ^{\alpha }\right) $
are taken to have odd parity.

For this considered theory the characteristic equations associated to the
gauge algebra (18, 19) becomes 
\begin{equation}
\delta V^{ij}-S(ij)\left[ -V^{kj}\frac{\delta (\delta \Phi ^{i})}{\delta
\Phi ^{k}}+Z^{\alpha j}\frac{\delta (\delta \Phi ^{i})}{\delta c^{\alpha }}%
\right] =0,  \tag{28-a}
\end{equation}
\begin{equation}
S(ijk)\left[ -V^{lk}\frac{\delta V^{ij}}{\delta \Phi ^{l}}+Z^{\alpha k}\frac{%
\delta V^{ij}}{\delta c^{\alpha }}+i\leftrightarrow j\right] =0,  \tag{28-b}
\end{equation}
\begin{equation}
\delta Z^{\alpha i}-\left[ -\frac{\delta (\delta c^{\alpha })}{\delta
c^{\beta }}+V^{ki}\frac{\delta (\delta c^{\alpha })}{\delta \Phi ^{k}}%
-Z^{\alpha k}\frac{\delta (\delta \Phi ^{i})}{\delta \Phi ^{k}}\right] =0, 
\tag{29-a}
\end{equation}
\begin{equation}
S\left( jk\right) \left[ -Z^{\beta j}\frac{\delta Z^{\alpha k}}{\delta
c^{\beta }}+V^{ij}\frac{\delta Z^{\alpha k}}{\delta \Phi ^{i}}\right] =0, 
\tag{29-b}
\end{equation}
where $S(...)$ means that a symmetrization over the indices in brackets is
carried out.

Let us now introduce the space $\complement $ of the $(d\times d)$
invertible matrices. One can define on $\complement $ (of dimension $d^{2}$)
a basis of $d^{2}$ matrices 
\begin{equation}
\{\Gamma ^{A}\}_{A=1,...,d^{2}},  \tag{30}
\end{equation}
which satisfies the orthonormality condition 
\begin{equation}
tr(\Gamma ^{A}\Gamma ^{B})=d\delta ^{AB},  \tag{31}
\end{equation}
where the trace operation is considered as the scalar product on the matrix
space. One may also define the inverse basis of (30) $\{\bar{\Gamma}%
^{A}\}_{A=1,...d^{2}}$ satisfying 
\begin{equation}
\bar{\Gamma}_{\alpha \lambda }^{A}\Gamma _{\lambda \beta }^{B}=\Gamma
_{\alpha \lambda }^{A}\bar{\Gamma}_{\lambda \beta }^{B}=\delta ^{AB}\delta
_{\alpha \beta }.  \tag{32}
\end{equation}

Furthermore, each matrix $M$ belonging to $\complement $ may be also
decomposed into a symmetric matrix and an antisymmetric one, i.e. $M_{\alpha
\beta }=M_{(\alpha \beta )}+M_{[\alpha \beta ]}$ where $M_{(\alpha \beta )}=%
\frac{1}{2}\left[ M_{\alpha \beta }+M_{\beta \alpha }\right] $ and $%
M_{[\alpha \beta ]}=\frac{1}{2}\left[ M_{\alpha \beta }-M_{\beta \alpha }%
\right] $. In other terms this means that $\complement $ can be decomposed
into two subspaces, i.e. $\complement =\complement _{0}\oplus \complement
_{1}$ where $\complement _{0}$ is the subspace of the symmetric matrices of
dimension $d(d+1)/2$ and $\complement _{1}$ is the subspace of the
antisymmetric matrices of dimension $d(d-1)/2$. From all the possible basis
on $\complement $, we will choose the one which is build from the basis of $%
\complement _{0}$ and $\complement _{1}$, in order to have 
\begin{equation}
\left( \Gamma ^{A}\right) ^{T}{}=(-)^{A}\Gamma ^{A},  \tag{33}
\end{equation}
where $A=0$ $(=1)$ for the $\Gamma ^{A}$ belonging to $\complement _{0}$ $%
(\complement _{1})$. Let now show that the introduction of such a basis for $%
\complement $ is of great help in the introduction of auxiliary fields and
then in performing the off-shell quantization of the theory. To this end,
one can put the full quantum action of the theory (25) in the form 
\begin{equation}
S_{q}=S+\frac{1}{4}V_{\alpha \beta }^{ij}F_{\rho ,i}F_{\sigma ,j}c^{\alpha
}c^{\beta }\bar{c}^{\rho }\bar{c}^{\sigma }+Q\Psi ,  \tag{34}
\end{equation}
where $F_{\rho ,i}=\delta F_{\rho }/\delta \Phi ^{i}$. We will focus us on
the second part of the right hand side of (34) 
\begin{equation}
\tilde{S}_{\Lambda }=\frac{1}{4}V_{\alpha \beta }^{ij}F_{\rho ,i}F_{\sigma
,j}c^{\alpha }c^{\beta }\bar{c}^{\rho }\bar{c}^{\sigma }.  \tag{35}
\end{equation}
By noting $F_{\rho ,i}F_{\sigma ,j}=F_{\rho \sigma ,ij}$, we can perform a
kind of Fierzing \cite{3} on (35). This is based on the observation that the
term $V_{\alpha \beta }^{ij}F_{\rho \sigma ,ij}$ can be viewed for fixed $%
\alpha $ and $\sigma $ as an $d\times d$ matrix which can be expanded into
the complete set of $\Gamma ^{A}$, we have 
\begin{equation}
V_{\alpha \beta }^{ij}F_{\rho \sigma ,ij}=C_{\alpha \sigma }^{A}\Gamma
_{\beta \rho }^{A},  \tag{36}
\end{equation}
where all the $C_{\alpha \sigma }^{A}$ are completely determined by (31) 
\begin{equation}
C_{\alpha \sigma }^{A}=\frac{(-)^{A}}{d}V_{\alpha \lambda }^{ij}\Gamma
_{\lambda \delta }^{A}F_{\delta \sigma ,ij}.  \tag{37}
\end{equation}
Doing the same operation once again on $V_{\alpha \lambda }^{ij}F_{\delta
\sigma ,ij}$ in (37), the action (35) can be cast in the form 
\begin{equation}
\tilde{S}_{\Lambda }=\frac{(-)^{B}}{4d^{2}}F_{\delta ,i}\Gamma _{\delta
\lambda }^{B}V_{\lambda \tau }^{ij}\Gamma _{\tau \gamma }^{A}F_{\gamma ,j}(%
\bar{c}^{\alpha }\Gamma _{\alpha \beta }^{A}c^{\beta })(\bar{c}^{\rho
}\Gamma _{\rho \sigma }^{B}c^{\sigma }).  \tag{38}
\end{equation}
We are now able to make the following identifications for the auxiliary
fields 
\begin{equation}
P^{A}\equiv (\bar{c}^{\alpha }\Gamma _{\alpha \beta }^{A}c^{\beta }). 
\tag{39}
\end{equation}
These fields have even parity and ghosts number zero. The action (38) will
then take the form 
\begin{equation}
\tilde{S}_{\Lambda }=W^{BA}P^{A}P^{B},  \tag{40}
\end{equation}
where 
\begin{equation}
W^{BA}=\frac{(-)^{B}}{4d^{2}}F_{,i}\Gamma ^{B}V^{ij}\Gamma ^{A}F_{,j}. 
\tag{41}
\end{equation}
By a direct calculation one finds: $F_{,i}\Gamma ^{B}V^{ij}\Gamma
^{A}F_{,j}=(-)^{A+B}F_{,i}\Gamma ^{A}V^{ij}\Gamma ^{B}F_{,j}$, so that $%
W^{BA}=W^{AB}$, and then no symmetrization is required in (40).

Since no ghost terms are explicitly occurring in the action (40) obtained
for the $d^{2}$ fields $P^{A}$, it can also be considered at the classical
level in the way that classically, we can put 
\begin{equation}
\tilde{S}=S+W^{AB}P^{A}P^{B},  \tag{42}
\end{equation}
which will represent the classical extension of the classical action $S$ of
the theory. Before investigating the symmetries of this action, an important
remark must be pointed out in order to show that the fields $P^{A}$ play
totally the role of auxiliary fields of the theory. The fact that the
classical extension (42) is algebraic in $P^{A}$ (it contains no derivative
terms in $P^{A}$) allows us to see that they are non propagating (non
dynamical) fields. They must also not introduce any new degrees of freedom
to the classical theory, i.e. their equations of motion derived from (42)
must be completely solved. This is simply guaranteed by the implicit
functions theorem \cite{9}. Indeed, at the dynamical level, the equations of
motion of the $d^{2}$ fields $P^{A}$ reads 
\begin{equation}
\frac{\delta \tilde{S}(\Phi ,P)}{\delta P^{A}}=0,  \tag{43}
\end{equation}
and the above mentioned theorem affirms that the condition 
\begin{equation}
\det \frac{\delta ^{2}\tilde{S}(\Phi ,P)}{\delta P^{A}\delta P^{B}}\neq 0, 
\tag{44}
\end{equation}
ensures that the system of the $d^{2}$ equations defined by (43) possesses a
unique system of $d^{2}$ solutions $P_{0}^{A}\left( \Phi _{0}^{i}\right) $,
where $\Phi _{0}^{i}$ are the solutions of the $N$ equations of motion of
the classical fields $\Phi ^{i}$, i.e., $\left( \delta \tilde{S}(\Phi
,P)/\delta \Phi ^{i}\right) _{\Phi _{0}^{i}}=0$. The condition (44) must be
viewed as crucial to check if any given classical theory can admit a
structure of auxiliary fields.

In view of (42), for any open gauge theory of type $(2,2)$ the condition
(44) leads to the fact that $W^{AB}$ must have an inverse $\bar{W}^{AB}$
such that 
\begin{eqnarray}
\bar{W}^{AB}W^{BC} &=&\delta ^{AC},  \TCItag{45-a} \\
W^{AB}\bar{W}^{BC} &=&\delta ^{AC}.  \TCItag{45-b}
\end{eqnarray}
Let us remark that these two conditions lead for $\bar{W}^{AB}$ as $W^{AB}$
to the same symmetry property.

Now, one can show that the action $\tilde{S}=S+W^{AB}P^{A}P^{B}$ is
invariant under the action of the operator $\Delta $ defined by 
\begin{eqnarray}
\Delta \Phi ^{i} &=&R_{\alpha }^{i}c^{\alpha }+K_{\alpha }^{iA}c^{\alpha
}P^{A}\text{ ,\ \ \ \ \ \ \ \ \ \ \ }  \TCItag{46-a} \\
\Delta P^{A} &=&L_{\alpha }^{iA}c^{\alpha }\frac{\delta \tilde{S}}{\delta
\Phi ^{i}}+E_{\alpha }^{AB}c^{\alpha }P^{B},  \TCItag{46-b}
\end{eqnarray}
where 
\begin{eqnarray}
K_{\alpha }^{iA}\left[ \Phi \right] &=&-\frac{1}{2d}V_{\alpha \beta
}^{ij}\Gamma _{\beta \lambda }^{A}F_{\lambda ,j}\text{ ,\ \ \ \ } 
\TCItag{47-a} \\
L_{\alpha }^{iA}\left[ \Phi \right] &=&-\frac{1}{2}\bar{W}^{AB}K_{\alpha
}^{iB},  \TCItag{47-b} \\
E_{\alpha }^{AB}\left[ \Phi \right] &=&-\frac{1}{2}\bar{W}^{AC}\frac{\delta
W^{CB}}{\delta \Phi ^{l}}R_{\alpha }^{l}.  \TCItag{47-c}
\end{eqnarray}
One may note that the explicit form of $K_{\alpha }^{iA}\left[ \Phi \right] $
(47-a) which extends the classical symmetry in (46-a) can be simply derived
by performing rearrangement of type (36) in the on-shell BRST transformation 
$Q\Phi ^{i}$ on the term $V_{\alpha \beta }^{ij}F_{\lambda ,j}$ viewed for
fixed $\alpha $.

The rest of our task is basically twofold. On the one hand we have to check
the $\Delta $-invariance of the full quantum action 
\begin{equation}
\tilde{S}_{q}=S+W^{AB}P^{A}P^{B}+\Delta \Psi ,  \tag{48}
\end{equation}
which contains the gauge-fixing terms. On the other hand, one has to show
that the defined BRST operator $\Delta $ is nilpotent off shell in order to
achieve the proof that the above introduced fields $P^{A}$ are the desired
auxiliary fields. However, one can remark that in view of (48) together with
the $\Delta $-invariance of $\tilde{S}$, the $\Delta $-invariance of $\tilde{%
S}_{q}$ simply requires that $\Delta ^{2}\Psi =0$ which is equivalent to
show the off-shell nilpotency of $\Delta $ on the classical fields $\Phi
^{i} $, and this because of the exclusive dependence on $\Phi ^{i}$ of the
gauge-fixing functions (20) for irreducible open gauge theories. To this
end, one has to add to the definition of $\Delta $ (46-a,b) and (47-a,c) its
action on the ghost fields 
\begin{equation}
\Delta c^{\lambda }=-\frac{1}{2}T_{\alpha \beta }^{\lambda }c^{\alpha
}c^{\beta }+H_{\alpha \beta }^{\lambda A}c^{\alpha }c^{\beta }P^{A}, 
\tag{49-a}
\end{equation}
where 
\begin{equation}
H_{\alpha \beta }^{\lambda A}\left[ \Phi \right] =\frac{1}{3d}Z_{\alpha
\beta \gamma }^{\lambda j}\Gamma _{\gamma \delta }^{A}F_{\delta ,j}, 
\tag{49-b}
\end{equation}
where the functions $Z_{\alpha \beta \gamma }^{\lambda j}$ acting on the
ghosts $c^{\alpha }c^{\beta }c^{\gamma }$ realize the non closure functions $%
Z^{\lambda j}$ defined in (14), i.e. $Z^{\lambda j}=\frac{1}{3}Z_{\alpha
\beta \gamma }^{\lambda j}c^{\alpha }c^{\beta }c^{\gamma }$. This leads to
the $\Delta $-invariance of $\tilde{S}_{q}$%
\begin{equation}
\Delta \tilde{S}_{q}=0.  \tag{50}
\end{equation}
We note, in particular, that to prove this we have used beside the
characteristic equations (28-a,b) the trivial but very helpful identity 
\begin{equation}
W^{AB}=-\frac{1}{2d}F_{,i}(\Gamma ^{A})^{T}K^{iB}.  \tag{51}
\end{equation}

We turn now to show the off-shell nilpotency of the BRST operator $\Delta $.
On the classical fields $\Phi ^{i}$ it is simply derived from (50) which
implies $\Delta ^{2}\Psi =0$, and then 
\begin{equation}
\Delta ^{2}\Phi ^{i}F_{\lambda ,i}=0.  \tag{52}
\end{equation}
On this ground, a particular observation on the gauge-fixing functions can
be done. These functions $F_{\lambda }\left[ \Phi \right] $ must not possess
any invariance whatever was the transformation on $\Phi ^{i}$, i.e. for any
set of transformations $\Delta _{\omega }\Phi ^{i}\equiv X_{\omega }^{i}%
\left[ \Phi \right] $, we must have 
\begin{equation}
\Delta _{\omega }F_{\lambda }=0\Rightarrow X_{\omega }^{i}=0,  \tag{53}
\end{equation}
where $``\omega "$ label the set of transformations of $\Phi ^{i}$. This
clearly leads to 
\begin{equation}
\forall X_{\omega }^{i}\left[ \Phi \right] :\text{ }X_{\omega
}^{i}F_{\lambda ,i}=0\Rightarrow X_{\omega }^{i}=0.  \tag{54}
\end{equation}
This condition on the gauge fixing functions allows us from (52) to prove
the off-shell nilpotency of $\Delta $ on the classical fields $\Phi ^{i}$.
That condition remains essential if we undertake to show this off-shell
nilpotency by a direct computation of $\Delta ^{2}\Phi ^{i}$. Indeed, it
permits us to obtain 
\begin{equation}
K_{\alpha }^{iA}\bar{W}^{AB}K_{\beta }^{jB}=V_{\alpha \beta }^{ij}\text{ }, 
\tag{55}
\end{equation}
which is necessary to the direct proof of 
\begin{equation}
\Delta ^{2}\Phi ^{i}=0.  \tag{56}
\end{equation}
Let us precise that in deriving (55) we have used the other trivial but
useful identity $\left[ K_{\alpha }^{iA}\bar{W}^{AB}K_{\beta
}^{jB}-V_{\alpha \beta }^{ij}\right] \Gamma _{\beta \sigma }^{D}F_{\sigma
,i}=0$ together with the condition (53) and the inverse basis of the $\Gamma
^{A}$ matrices.

We have now to show the off-shell nilpotency of $\Delta $ on the ghost
fields $c^{\alpha }$. To this end, beside the characteristic equations
(29-a,b) we use 
\begin{equation}
Z_{\alpha \beta \gamma }^{\lambda j}=-H_{\alpha \beta }^{\lambda A}L_{\gamma
}^{jA},  \tag{57}
\end{equation}
which is easily proven from the identity $\left[ Z_{\alpha \beta \gamma
}^{\lambda j}+H_{\alpha \beta }^{\lambda A}L_{\gamma }^{jA}\right] \Gamma
_{\rho \sigma }^{B}F_{\sigma }^{j}=0$ in the same way that we have done for
Eq. (55). Therefore, we find 
\begin{equation}
\Delta ^{2}c^{\alpha }=0.  \tag{58}
\end{equation}

Finally, the off-shell nilpotency of $\Delta $ on the auxiliary fields $%
P^{A} $ can be simply deduced from (56) and (58). Indeed, the evaluation of $%
\Delta ^{3}\Phi ^{i}$ $=\Delta (\Delta ^{2}\Phi ^{i})=\Delta ^{2}(\Delta
\Phi ^{i})$ leads to 
\begin{equation}
\Delta ^{2}\Phi ^{k}\frac{\delta (\Delta \Phi ^{i})}{\delta \Phi ^{k}}%
+\Delta ^{2}c^{\alpha }\frac{\delta (\Delta \Phi ^{i})}{\delta c^{\alpha }}%
+\Delta ^{2}P^{A}\frac{\delta (\Delta \Phi ^{i})}{\delta P^{A}}=0,  \tag{59}
\end{equation}
which, in view of (56), (58) and (46-a) implies 
\begin{equation}
\Delta ^{2}P^{A}K_{\alpha }^{iA}=0,  \tag{60}
\end{equation}
then, by the application of $\Gamma _{\alpha \beta }^{B}F_{\beta ,i}$, which
is non-degenerate in view of (53) and the existence of $\bar{\Gamma}^{A}$,
it follows that 
\begin{equation}
W^{AB}\Delta ^{2}P^{A}=0.  \tag{61}
\end{equation}
Using the fact that an inverse for $W^{AB}$ must exist, we get 
\begin{equation}
\Delta ^{2}P^{A}=0,  \tag{62}
\end{equation}
which ends up with the proof that the BRST operator $\Delta $ given by the
above prescription applied for open gauge theories of type $(2,2)$ is
nilpotent off shell.

\subsection{Open gauge theories of type $(p,q)$}

Although the general case of open gauge theories of type $(p,q)$ contains
more characteristic gauge functions as well as more associated
characteristic equations (18, 19), almost of all the general features
leading to build up the off-shell version of an on-shell open gauge theory
are expressed in the case of theories of type $(2,2)$. Indeed, the typical
rearrangement introduced in (36) together with the field redefinition (39)
which allow us to identify the auxiliary fields of the theory and the
crucial condition (44) remains unchanged and sufficient to formally find out
the off-shell BRST operator and the classical extension for any given open
gauge theory of gauge fields $\Phi ^{i}$ enriched with the set of auxiliary
fields. We then only concentrate on particular remarks that stand out in the
general case, all other results will be directly given. These remarks are
basically twofold. The first one affects the general form of the action
obtained for the on-shell quantum theory (25). This action contains clearly
higher order ghost-antighost couplings and could be recast in the form 
\begin{equation}
S_{q}=S+Q\Psi -\sum_{n=1}^{p-1}\frac{n}{n+1}\delta _{n}\Psi \text{ }, 
\tag{63}
\end{equation}
where $Q$ is the on-shell BRST operator defined by (24-a,b). Expressing each
term of $\sum_{n=1}^{p-1}\frac{n}{n+1}\delta _{n}\Psi $ occurring in the
above expression by using (22-b) one obtains 
\begin{equation}
\frac{n}{n+1}\delta _{n}\Psi =\frac{(-)^{n}n}{(n+1)!}%
(-)^{a_{n+1}}V_{n+1}^{i_{1}...i_{n+1}}\Psi _{,i_{1}}...\Psi _{,i_{n+1}}\text{
,}  \tag{64}
\end{equation}
developing then the $V_{n}^{i_{1}...i_{n}}$ functions in terms of the ghost
fields in the same way we have done in (11), 
\begin{equation}
V_{n}^{i_{1}...i_{n}}=\frac{1}{n}(-)^{\sum_{s=1}^{n-1}(\alpha
_{s}+1)\sum_{r=1}^{n}\alpha _{r}}(-)^{\sum_{r,s=1}^{n}(i_{r}\alpha
_{s})}V_{\alpha _{1}...\alpha _{n}}^{i_{1}...i_{n}}\left[ \Phi \right]
c^{\alpha _{1}}...c^{\alpha _{n}}\text{ ,}  \tag{65}
\end{equation}
and also expressing the gauge fermion $\Psi $ in function of the antighost
fields, using $\Psi =F_{\beta }\left[ \Phi \right] \bar{c}^{\beta }$, we
find that for any order $``n"$ in (63) a term of type $V_{\alpha
_{1}...\alpha _{n+1}}^{i_{1}...i_{n+1}}F_{\beta _{1},i_{1}}...F_{\beta
_{n+1},i_{n+1}}c^{\alpha _{1}}...c^{\alpha _{n+1}}\bar{c}^{\beta _{1}}...%
\bar{c}^{\beta _{n+1}}$ contributes to the quantum action. They are of even
order in ghost-antighost pairs whatever the integer $``n"$ is. By performing
the Fierz-like rearrangement (see Eq.(36)) $n+1$ times on each coefficient
of these terms using the orthonormality property of the basis $\{\Gamma
^{A}\}_{A=1,...d}$ (31) we show that they can be put in the form 
\begin{eqnarray}
V_{\alpha _{1}...\alpha _{n+1}}^{i_{1}...i_{n+1}}F_{\beta
_{1},i_{1}}...F_{\beta _{n+1},i_{n+1}} &=&\frac{1}{d^{n+1}}V_{\rho
_{1}...\rho _{n+1}}^{i_{1}...i_{n+1}}F_{\sigma _{1},i_{1}}...F_{\sigma
_{n+1},i_{n+1}}\Gamma _{\rho _{1}\sigma _{1}}^{A_{1}}...\Gamma _{\rho
_{n+1}\sigma _{n+1}}^{A_{n+1}}  \notag \\
&&\times \Gamma _{\beta _{1}\alpha _{1}}^{A_{1}}...\Gamma _{\beta
_{n+1}\alpha _{n+1}}^{A_{n+1}}\text{ },  \TCItag{66}
\end{eqnarray}
where a sum over $(A_{1},...A_{n+1})$ is underlaid. Thus the higher order
terms in the quantum action (63) acquire the form 
\begin{equation}
\tilde{S}_{\Lambda }=\sum_{n=1}^{p-1}W_{n+1}^{A_{1}...A_{n+1}}\left[ \Phi %
\right] (\bar{c}\Gamma ^{A_{1}}c)...(\bar{c}\Gamma ^{A_{n+1}}c)\text{ ,} 
\tag{67}
\end{equation}
where all the coefficients $W_{n+1}^{A_{1}...A_{n+1}}\left[ \Phi \right] $
are completely defined by Eqs. (64), (65) and (66). We are now able to
perform the same identifications as in the previous subsection for the
auxiliary fields (39), i.e. $P^{A}\equiv \left( \bar{c}\Gamma ^{A}c\right) $%
. To step forward we have to make an other remark which can be crucial for
practical application of our prescription. In the general case the $d^{2}$
fields $P^{A}$ constructed in this way have no defined grassmannian parity.
Indeed, since the ghost and antighost fields $(\bar{c}^{\alpha },c^{\beta })$
associated to the classical symmetry parameters $(\varepsilon ^{\alpha })$
have various grassmannian parities, any bilinear combination of them will
not have any defined parity. For that reason this redefinition is taken to
be purely formal. For practical application we have to split the formal set
of fields $P^{A}$ into sets having well defined parities. This can easily
done in the following way. The general set of the $``d"$ symmetries can be
divided into the set of the $``d_{b}"$ bosonic symmetries and the set of the 
$``d_{f}"$ fermionic\textit{\ }ones, such that $d=d_{b}+d_{f}$. Then, each $%
\Gamma ^{A}$ of the $d^{2}$ elements of the basis of the $d\times d$ matrix
space $\complement $ can take the following bloc matrix form 
\begin{equation}
\Gamma _{d\times d}^{A}\equiv \left[ 
\begin{array}{ll}
\Gamma _{d_{b}\times d_{b}}^{A^{1}} & \Gamma _{d_{f}\times d_{b}}^{A^{2}} \\ 
\Gamma _{d_{b}\times d_{f}}^{A^{3}} & \Gamma _{d_{f}\times d_{f}}^{A^{4}}
\end{array}
\right] ,  \tag{68}
\end{equation}
which can be condensed in the notation $\Gamma ^{A}\equiv \left( \Gamma
^{A^{a}}\right) _{a=1,...,4}$, where each value of $``a"$ denotes one of the
four sectors of $\Gamma ^{A}$. Then the set of the fields $P^{A}$ can be
viewed as a supermultiplet\textit{\ }containing the bosonic as well as the
fermionic auxiliary fields, i.e. $P^{A}\equiv \left( P^{A^{a}}\right)
_{a=1,...,4}$, where every auxiliary field $P^{A^{a}}$ is introduced by the
field redefinition $P^{A^{a}}\equiv \left( \bar{c}\Gamma ^{A^{a}}c\right) $.
The $d_{b}^{2}$ fields $P^{A^{1}}$ and the $d_{f}^{2}$ fields $P^{A^{4}}$
are bosonic while the $d_{f}\times d_{b}$ fields $P^{A^{2}}$ and the $%
d_{b}\times d_{f}$ fields $P^{A^{3}}$ are fermionic. All of them are of
ghost number zero. Then the action (67) which formally reads $\tilde{S}%
_{\Lambda }=\sum_{n=1}^{p-1}W_{n+1}^{A_{1}...A_{n+1}}\left[ \Phi \right]
P^{A_{1}}...P^{A_{n+1}}$, will be practically written as 
\begin{equation}
\tilde{S}_{\Lambda
}=\sum_{n=1}^{p-1}%
\sum_{a_{1}...a_{n+1}=1}^{4}W_{n+1}^{A_{1}^{a_{1}}...A_{n+1}^{a_{n+1}}}\left[
\Phi \right] P^{A_{1}^{a_{1}}}...P^{A_{n+1}^{a_{n+1}}},  \tag{69}
\end{equation}
where the functions $W_{n+1}^{A_{1}^{a_{1}}...A_{n+1}^{a_{n+1}}}\left[ \Phi 
\right] $ are completely derived upon the $\Gamma ^{A}$-dependence of $%
W_{n+1}^{A_{1}...A_{n+1}}\left[ \Phi \right] $, see Eq.(66) and the
definition of the $P^{A^{a}}$.

In what follows we pursue only with the formal notation $P^{A}$ for the
auxiliary fields, but keeping in mind that for practical applications we
have to go back to the fields $P^{A^{a}}$ in order to obtain the correct
representation of the auxiliary fields.

Let us now introduce the classical extension $\tilde{S}\left( \Phi ,P\right) 
$ of the classical action of the theory $S\left( \Phi \right) $%
\begin{equation}
\tilde{S}\left( \Phi ,P\right) =S\left( \Phi \right)
+\sum_{n=1}^{p-1}W_{n+1}^{A_{1}...A_{n+1}}\left[ \Phi \right]
P^{A_{1}}...P^{A_{n+1}},  \tag{70}
\end{equation}
and by applying the same procedure as for the $(2,2)$-type open gauge
theories, one expands in ghost-antighost pairs the on-shell BRST operator $Q$
acting on the gauge fields $\Phi ^{i}$ (24-a) in order to obtain the
off-shell BRST symmetry of the classical action (70). Each term of $Q$, i.e. 
$\delta _{n}\Phi ^{i}=\frac{1}{n!}(-1)^{in+a_{n}}V_{n+1}^{ii_{1}...i_{n}}%
\Psi _{,i_{1}}...\Psi _{,i_{n}}$ clearly contains $``n"$ pairs $(\bar{c}%
^{\alpha },c^{\beta })$, then by performing $``n"$ times the Fierz-like
rearrangement and also make the suitable identification for the auxiliary
fields $P^{A}$ we obtain the following BRST transformation on $\Phi ^{i}$%
\begin{equation}
\Delta \Phi ^{i}=\delta \Phi ^{i}+\sum_{n=1}^{p-1}K_{n\alpha
}^{iA_{1}...A_{n}}\left[ \Phi \right] c^{\alpha }P^{A_{1}}...P^{A_{n}}, 
\tag{71}
\end{equation}
where $\delta $ is the standard BRST operator.

In order to consider the fields $P^{A}$ as auxiliary fields we still impose
the general condition $\det \delta ^{2}\tilde{S}(\Phi ,P)/\delta P^{A}\delta
P^{B}\neq 0$. To this purpose it is convenient to put the action (70) in the
form 
\begin{equation}
\tilde{S}\left( \Phi ,P\right) =S+\hat{W}^{AB}\left[ \Phi ,P\right]
P^{A}P^{B},  \tag{72}
\end{equation}
where $\hat{W}^{AB}\left[ \Phi ,P\right] =W_{2}^{AB}\left[ \Phi \right]
+\sum_{n=3}^{p-1}W_{n}^{ABC_{1}...C_{n-2}}\left[ \Phi \right]
P^{C_{1}}...P^{C_{n-2}}$, then the condition (44) will just imply that $\hat{%
W}^{AB}\left[ \Phi ,P\right] $ must have an inverse $\hat{W}_{inv}^{AB}\left[
\Phi ,P\right] $ such that $\hat{W}_{inv}^{AB}\hat{W}^{BC}=\delta ^{AC}$ and 
$\hat{W}^{AB}\hat{W}_{inv}^{BC}=\delta ^{AC}$. In the same way the BRST
transformation (71) could be cast in the form 
\begin{equation}
\Delta \Phi ^{i}=\delta \Phi ^{i}+\hat{K}_{\alpha }^{iA}\left[ \Phi ,P\right]
c^{\alpha }P^{A},  \tag{73}
\end{equation}
where $\hat{K}_{\alpha }^{iA}\left[ \Phi ,P\right] =K_{1\alpha }^{iA}\left[
\Phi \right] +\sum_{n=2}^{p-1}K_{n\alpha }^{iAA_{1}...A_{n-1}}\left[ \Phi %
\right] P^{A_{1}}...P^{A_{n-1}}$. Then by defining the action of $\Delta $
on the auxiliary fields 
\begin{equation}
\Delta P^{A}=-\frac{1}{2}\hat{W}_{inv}^{AB}\hat{K}_{\alpha }^{iB}c^{\alpha }%
\tilde{S}_{,i}-\hat{W}_{inv}^{AC}\delta \hat{W}^{AB},  \tag{74}
\end{equation}
a tedious but a straightforward calculation leads to the $\Delta $%
-invariance of the classical extension $\tilde{S}\left( \Phi ,P\right) $.

The last step will consist in showing the off-shell nilpotency of the BRST
operator $\Delta $. To this purpose we supplement the definition of $\Delta $
with its application on the ghost fields $c^{\alpha }$ in the same spirit as
in the case of the gauge fields $\Phi ^{i}$. First we begin to expand in
ghost-antighost pairs the on-shell BRST operator $Q$ acting on $c^{\alpha }$
(24-a), this involves functions of type $Z_{\lambda _{1}...\lambda
_{n+2}}^{\alpha i_{1}...i_{n}}$ which realize the characteristic functions
of type $Z_{n+1}^{\alpha i_{1}...i_{n}}$ (see Eq.(23-b)) by acting on the $%
(n+2)-th$ $order$ term $c^{\lambda _{1}}...c^{\lambda _{n+2}}$ as well as
the gauge fixing terms $F_{\beta _{1},i_{1}}...F_{\beta _{n+1},i_{n+1}}$
related with the $n-th$ $order$ term in antighost fields $\bar{c}^{\beta
_{1}}...\bar{c}^{\beta _{n+1}}$. Thus, each term in the definition of the
on-shell BRST operator $Q$ contributes with a term of order $``n"$ in
ghost-antighost pairs $(\bar{c}^{\alpha },c^{\beta })$, then performing $%
``n" $ times the Fierz-like rearrangement (36) and also applying the
prescribed identification for the auxiliary fields $P^{A}$ we obtain the
following form for the BRST transformation on $c^{\alpha }$ (for best
insight, one may return to Eqs. (49-a,b)) 
\begin{equation}
\Delta c^{\alpha }=\delta c^{\alpha }+\sum_{n=1}^{q-1}H_{n\rho \sigma
}^{\alpha A_{1}...A_{n}}\left[ \Phi \right] c^{\rho }c^{\sigma
}P^{A_{1}}...P^{A_{n}},  \tag{75}
\end{equation}
which can be easily put in the more convenient expression 
\begin{equation}
\Delta c^{\alpha }=\delta c^{\alpha }+\hat{H}_{\rho \sigma }^{\alpha A}\left[
\Phi ,P\right] c^{\rho }c^{\sigma }P^{A},  \tag{76}
\end{equation}
where $\hat{H}_{\rho \sigma }^{\alpha A}\left[ \Phi ,P\right] =H_{1\rho
\sigma }^{\alpha A}\left[ \Phi \right] +\sum_{n=2}^{q-1}H_{n\rho \sigma
}^{\alpha A_{1}...A_{n-1}}\left[ \Phi \right] c^{\rho }c^{\sigma
}P^{A_{1}}...P^{A_{n-1}}$. Then we can show by a last tedious calculation,
that the obtained BRST operator $\Delta $ defined by Eqs. (73), (74) and
(75) is nilpotent off shell, i.e., 
\begin{equation}
\Delta ^{2}X=0,  \tag{77}
\end{equation}
where $X$ describes all the fields of the theory. However, let us note that
in addition to the characteristic equations (18 and 19) the proof of the
off-shell nilpotency of $\Delta $ requires the condition (53) imposed on the
gauge fixing functions.

Once we get the off-shell nilpotency of $\Delta $, the gauge fixing action
occurring in the full quantum action of the theory can be put in the usual $%
\Delta $-exact form, i.e., $S_{q}=\tilde{S}+\Delta \Psi $.

\section{\label{IV}Minimal and non-minimal set of auxiliary fields}

We are going now to investigate one of the most typical feature of theories
that contain auxiliary fields. For those theories we remark that the number
of auxiliary fields is not unique, but in all cases we may find a minimal
set of these fields (for a review see Refs. [3] and [10]). In this chapter
we will see how this statement can be analyzed and reproduced in the general
framework of the ideas suggested in this paper. We firstly deal with
theories of type $(2,2)$ then we briefly discuss the general case $(p,q)$
which doesn't bring anything new to the spirit of the approach.

In the above chapters we show how we can start with an on-shell open gauge
theory to end up with the corresponding off-shell version. The procedure is
essentially based on the identification (39) for the auxiliary fields, i.e., 
\begin{equation}
P^{A}\equiv (\bar{c}^{\alpha }\Gamma _{\alpha \beta }^{A}c^{\beta }), 
\tag{78}
\end{equation}
which are clearly of number $``d^{2}"$. The set of the $``d^{2}"$ matrices $%
\{\Gamma ^{A}\}_{A=1,...,d^{2}}$ can be always split into the two sets of
the symmetric matrices $\{\Gamma _{0}^{A}\}$ of number $d(d+1)/2$ and the
antisymmetric matrices $\{\Gamma _{1}^{A}\}$ of number $d(d-1)/2$. This fact
together with the identification (78) permit us to split the set of
auxiliary fields noted $\Lambda _{p}$ into two parts. The first one $\Lambda
_{p}^{0}$ contains $d(d+1)/2$ auxiliary fields $P_{0}^{A}$ defined by 
\begin{equation}
P_{0}^{A}\equiv \bar{c}^{\alpha }\Gamma _{0\alpha \beta }^{A}c^{\beta }, 
\tag{79-a}
\end{equation}
and the second part $\Lambda _{p}^{1}$ contains $d(d-1)/2$ auxiliary fields $%
P_{1}^{A}$ defined by 
\begin{equation}
P_{1}^{A}\equiv \bar{c}^{\alpha }\Gamma _{1\alpha \beta }^{A}c^{\beta }. 
\tag{79-b}
\end{equation}

Our task consists now in showing that we could eliminate one of the two
above representations of auxiliary fields without affecting the other one.
To this aim one can remark that the auxiliary fields $P^{A}$ appears in the
off-shell version of the theory at two levels: in the classical extension of
the classical action (42) and in the off-shell BRST operator $\Delta $. In
both of them they are associated to coefficients that involve the
characteristic functions of the theory and the different gauge fixing
functions. It is the last dependence that will be investigated. We first
introduce from the gauge fixing functions $F_{\alpha }\left[ \Phi \right] $
a set of functions $F_{\alpha }^{A}\left[ \Phi \right] $ defined by 
\begin{equation}
F_{\alpha }^{A}\left[ \Phi \right] =\Gamma _{\alpha \beta }^{A}F_{\beta }%
\left[ \Phi \right] .  \tag{80}
\end{equation}
Such a definition is guaranteed by the existence of the inverse basis $\bar{%
\Gamma}^{A}$. Thus we have 
\begin{equation}
F_{\alpha }\left[ \Phi \right] =\bar{\Gamma}_{\alpha \beta }^{A}F_{\beta
}^{A}\left[ \Phi \right] .  \tag{81}
\end{equation}
We can observe that the inverse basis $\bar{\Gamma}^{A}$ can also be
decomposed into symmetric and antisymmetric parts $\bar{\Gamma}_{0}^{A}$ and 
$\bar{\Gamma}_{1}^{A}$ in the way that using (32) we obtain 
\begin{equation}
F_{\alpha }\left[ \Phi \right] =\bar{\Gamma}_{0\alpha \beta }^{A}F_{0\beta
}^{A}\left[ \Phi \right] +\bar{\Gamma}_{1\alpha \beta }^{A}F_{1\beta }^{A}%
\left[ \Phi \right] .  \tag{82}
\end{equation}
Upon this decomposition, the classical extension (42) reads 
\begin{equation}
\tilde{S}%
=S+W_{00}^{AB}P_{0}^{A}P_{0}^{B}+W_{11}^{AB}P_{1}^{A}P_{1}^{B}+W_{10}^{AB}P_{1}^{A}P_{0}^{B}+W_{01}^{AB}P_{0}^{A}P_{1}^{B},
\tag{83}
\end{equation}
with 
\begin{eqnarray}
W_{00}^{AB} &=&\frac{1}{4d^{2}}F_{0}^{Ai}V^{ij}F_{0}^{Bj},  \TCItag{84-a} \\
W_{11}^{AB} &=&\frac{1}{4d^{2}}F_{1}^{Ai}V^{ij}F_{1}^{Bj},  \TCItag{84-b} \\
W_{10}^{AB} &=&\frac{1}{4d^{2}}F_{1}^{Ai}V^{ij}F_{0}^{Bj},  \TCItag{84-c} \\
W_{01}^{AB} &=&\frac{1}{4d^{2}}F_{0}^{Ai}V^{ij}F_{1}^{Bj}.  \TCItag{84-d}
\end{eqnarray}
Note that $W_{00}$ and $W_{11}$ are symmetric in $A$ and $B$, and $%
W_{10}^{AB}=W_{01}^{BA}$. We are now able to choose between the elimination
of the fields $P_{0}^{A}$ or $P_{1}^{A}$. This will be simply done by taking
advantage of the freedom in the manner that we choose the gauge fixing
functions. If we want, for example, to eliminate the fields $P_{0}^{A}$ it
is sufficient to choose the gauge fixing functions such that in (82) we have 
\begin{equation}
F_{0\beta }^{A}=0.  \tag{85}
\end{equation}
From this and from (84-a,d), the only coefficient that remains in (83) is $%
W_{11}$ and only the auxiliary fields $P_{1}^{A}$ take part in the classical
extension of the action. In order to completely eliminate the $P_{0}^{A}$ it
is necessary to show that they do not appear into the BRST operator $\Delta $%
. Indeed, from (46-a,b) and (49) we find 
\begin{eqnarray}
\tilde{Q}\Phi ^{i} &=&R_{\alpha }^{i}c^{\alpha }+K_{1\alpha }^{iA}c^{\alpha
}P_{1}^{A}\text{,\ \ \ \ \ \ \ }  \TCItag{86-a} \\
\tilde{Q}C^{\lambda } &=&-\frac{1}{2}T_{\alpha \beta }^{\lambda }c^{\alpha
}c^{\beta }+H_{1\alpha \beta }^{\lambda A}c^{\alpha }c^{\beta }P_{1}^{A}, 
\TCItag{86-b} \\
\tilde{Q}P_{1}^{A} &=&L_{1\alpha }^{iA}c^{\alpha }\frac{\delta \tilde{S}_{0}%
}{\delta \Phi ^{i}}+E_{1\alpha }^{AB}c^{\alpha }P_{1}^{B},  \TCItag{86-c} \\
\tilde{Q}P_{0}^{A} &=&0,  \TCItag{86-d}
\end{eqnarray}
with 
\begin{eqnarray}
K_{1\alpha }^{iA}\left[ \Phi \right] &=&-\frac{1}{2d}V_{\alpha \beta
}^{ij}F_{1\beta }^{Aj}\text{ ,}  \TCItag{87-a} \\
H_{1\alpha \beta }^{\lambda A}\left[ \Phi \right] &=&\frac{1}{3!d}Z_{\alpha
\beta \gamma }^{\lambda j}F_{1\gamma }^{Aj}\text{\ ,}  \TCItag{87-b} \\
L_{1\alpha }^{iA}\left[ \Phi \right] &=&-\frac{1}{2}\bar{W}%
_{11}^{AB}K_{1\alpha }^{iB}\text{ },  \TCItag{87-c} \\
E_{1\alpha }^{AB}\left[ \Phi \right] &=&-\frac{1}{2}\bar{W}_{11}^{AC}\frac{%
\delta W_{11}^{CB}}{\delta \Phi ^{l}}R_{\alpha }^{l},  \TCItag{87-d}
\end{eqnarray}
where $\bar{W}_{11}$ is the inverse of $W_{11}$.

Thus the condition (85) is sufficient to the elimination of the auxiliary
fields $P_{0}^{A}$. Moreover, one can note that if instead of (85) we have
chosen the gauge fixing functions such that $F_{1\beta }^{A}=0$, then the
fields $P_{1}^{A}$ will be eliminated. So we have defined two possible
configurations for the auxiliary fields. For a given open gauge theory, the
choice of the gauge fixing functions such that $F_{0\beta }^{A}=0$ leads to
the set $\Lambda _{p}^{1}$ of the $d(d-1)/2$ auxiliary fields $P_{1}^{A}$.
This will be named \textit{the minimal set of auxiliary fields}. The other
choice of the gauge fixing functions such that $F_{1\beta }^{A}=0$ which
leads to the set $\Lambda _{p}^{0}$ of the $d(d+1)/2$ auxiliary fields $%
P_{0}^{A}$ will be named \textit{the non-minimal set of auxiliary fields}.

Since the keystone for the determination of the minimal (or non-minimal) set
of auxiliary fields is the choice of the gauge fixing functions via the
decomposition (82), no particular generalization is needed in the case of
theory of type $(p,q)$. The condition $F_{0\beta }^{A}=0$ $(F_{1\beta
}^{A}=0)$ remains sufficient to obtain the minimal (non-minimal) set of
auxiliary fields for general open gauge theories. Nevertheless, one can
recall that for a practical application (where both of bosonic and fermionic
symmetries are responsible for the opening of the classical algebra), we
have to deal with the set $\left( P^{A^{a}}\right) _{a=1,...,4}$ of the
genuine auxiliary fields with well defined parities obtained from the formal
set $\left( P^{A}\right) $ as it is shown in the second part of section \ref
{III}. In order to understand what will occur to the minimal and non-minimal
configurations of auxiliary fields, we must notice that the $%
``d^{2}=(d_{b}+d_{f})^{2}"$ matrices $\Gamma ^{A}$ expressed such as in (68)
lead to the $\left( d_{b}(d_{b}-1)/2+d_{f}(d_{f}-1)/2+d_{b}d_{f}\right) $
antisymmetric matrices of the base of $\complement _{1}$ and the $\left(
d_{b}(d_{b}+1)/2+d_{f}(d_{f}+1)/2+d_{b}d_{f}\right) $ symmetric matrices of
the base of $\complement _{0}$. Therefore the minimal set $\Lambda _{p}^{1}$
will contain $\left( d_{b}(d_{b}-1)/2+d_{f}(d_{f}-1)/2\right) $ bosonic and $%
\left( d_{b}d_{f}\right) $ fermionic auxiliary fields, while the non-minimal
set $\Lambda _{p}^{0}$ will contain $\left(
d_{b}(d_{b}+1)/2+d_{f}(d_{f}+1)/2\right) $ bosonic and $\left(
d_{b}d_{f}\right) $ fermionic auxiliary fields.

To end this chapter, we will briefly discuss the particular case of simple
supergravity (D=4 and N=1) to show how the procedure developed in this paper
can be practically applied. In this theory \cite{3} the classical dynamical
gauge fields are the vierbein $e_{\mu }^{a}$ and the gravitino $\psi _{\mu
}^{A}$ with $a=1,...,4$ label the flat Minkowski space, $\mu =1,...,4$ label
the curved Riemannian space and $A=1,...,4$ is related to the $N=1$
supersymmety. One recalls that the theory admits a vanishing torsion leading
to a non propagating spin connection $\omega _{\mu }^{ab}$. The symmetries
of the theory are the diffeomorphism, the Lorenz and the supersymmetry
transformations. Their associated ghost fields are $c^{\mu }$, $c^{ab}$ and $%
c^{A}$ respectively. The classical BRST operator associated to the classical
symmetries of the theory have the following on-shell property \cite[er]{3} 
\begin{eqnarray}
\delta ^{2}\psi _{\mu } &=&V_{\mu \nu }\frac{\delta S}{\delta \bar{\psi}%
_{\nu }},  \TCItag{88-a} \\
\delta ^{2}c^{ab} &=&Z_{\nu }^{ab}\frac{\delta S}{\delta \bar{\psi}_{\nu }},
\TCItag{88-b} \\
\delta ^{2}X &=&0\text{ \quad \textit{for all others fields,}}  \TCItag{88-c}
\end{eqnarray}
with $\bar{\psi}_{\nu }=\psi _{\nu }^{T}C$, where $C$ is the charge
conjugation matrix, and the supersymmetric index is omitted for simplicity.
This on-shell structure follows easily from the open structure of the
superalgebra of the simple supergravity. The characteristic functions $%
V_{\mu \nu }$ and $Z_{\nu }^{ab}$ are given by 
\begin{eqnarray}
V_{\mu \nu } &=&\frac{1}{8}\bar{c}\gamma ^{a}c\left( \frac{1}{4}g_{\mu \nu
}\gamma _{a}-\frac{1}{2}e\varepsilon _{\mu \nu \rho \tau }e_{b}^{\tau
}\gamma _{5}\gamma ^{b}\right)  \TCItag{89} \\
&&+\frac{1}{8}\bar{c}\sigma _{ab}c\left( e_{\mu }^{a}e_{\nu }^{b}+\frac{1}{2}%
g_{\mu \nu }\sigma ^{ab}-\frac{1}{2}e\varepsilon _{\mu \nu \rho \tau
}e_{b}^{\tau }\gamma _{5}-\frac{1}{2}e\varepsilon _{\mu \nu \rho \tau
}e^{\rho a}e^{\tau b}\gamma _{5}\right)  \notag
\end{eqnarray}
\begin{equation}
Z_{\mu }^{ab}=\frac{1}{8}\bar{c}\gamma _{a}e_{\mu }^{a}\sigma ^{ab}\gamma
_{5}c\bar{c}\gamma _{5},  \tag{90}
\end{equation}
where $e=det(e_{\mu }^{a})$ and $g_{\mu \nu }=e_{\mu }^{a}e_{a\mu }$. These
characteristic functions are related upon characteristic equations \cite{6}
of type (28-a,b) and (29-a,b) and show that simple supergravity is of type $%
(2,2)$. Since the only symmetry that is responsible for the opening of the
classical algebra is supersymmetry, and following the procedure presented in
this paper, the complete set of auxiliary fields will contain $%
d^{2}=4^{2}=16 $ bosonic fields. To step forward and find out the complete
representation of the auxiliary fields we need to define a convenient basis
for the $4\times 4$ matrix. Such a basis is given by the 16 matrix $\{\Gamma
^{A}\}_{A=1,...16}\equiv (C,$ $C\gamma ^{a},$ $2C\sigma ^{ab},$ $C\gamma
_{5}\gamma ^{a},$ $C\gamma _{5})$, where $\gamma ^{a}$ are the Dirac matrix, 
$\sigma ^{\mu \nu }=\frac{1}{4}\left[ \gamma ^{a},\gamma ^{b}\right] $ and $%
\gamma _{5}=\gamma _{1}\gamma _{2}\gamma _{3}\gamma _{4}$. By taking
advantage of the properties of the Dirac matrices one can show that this set
of matrices split into the set of the six antisymmetric matrices $(C,$ $%
C\gamma _{5},C\gamma _{5}\gamma ^{a})$ and the ten symmetric ones $(C\gamma
^{a},$ $2C\sigma ^{ab})$. According to this basis the sixteen bosonic
degrees of freedom expected for the auxiliary fields will be distributed
with respect to the following multiplet representation $\left(
S(scalar),P(pseudoscalar),A_{5}^{a}(pseudovector)\right) $ for the minimal
set and $\left( A^{a}(vector),E^{ab}(2nd-rank\text{ }antisymmetric\text{ }%
tensor)\right) $ for the non-minimal one. These are the standard results
occurring in simple supergravity. Let us note that once we choose the
standard gauge fixing function for supergravity i.e. $F=e\gamma ^{\mu }\psi
_{\mu }$ we can see that the only coefficient $W_{11}^{AB}$ (84-b) that
remains in the minimal representation of auxiliary fields acquires the
following simple form 
\begin{equation}
W_{11}=-\frac{e}{3}\left[ 
\begin{array}{ccc}
1 & 0 & 0 \\ 
0 & -1 & 0 \\ 
0 & 0 & g_{\mu \nu }
\end{array}
\right] ,  \tag{91}
\end{equation}
which arises from the particular property of the characteristic function $%
V_{\mu \nu }$, that for any arbitrary spinor $\varphi $ we have $\bar{c}%
\gamma _{\nu }V_{\mu \rho }\gamma ^{\rho }\varphi =0$. This directly leads,
from (83), to the usual classical extension 
\begin{equation}
\tilde{S}=S_{cl}-\frac{e}{3}(S^{2}-P^{2}+A_{a}A^{a}).  \tag{92}
\end{equation}
One can also easily derive the associated BRST symmetry which is nilpotent
off shell from the general equations (86-a,d) and (87-a,d) and find the
standard results (see Refs. [3] and [6]).

\section{\label{V}Concluding Remarks}

In this paper we have presented a prescription leading to the construction
of an off-shell BRST quantization scheme for irreducible open gauge
theories. We first obtained the on-shell BRST full quantum action together
with its associated on-shell BRST symmetries. This is realized upon taking
advantage of the characteristic functions related to corresponding equations
that characterize general open gauge algebras. From this follows the
construction of the off-shell version of the theory. To this aim, we used a
suitable field redefinition which permits us to find out the necessary set
of auxiliary fields which leads to the classical extension of the classical
action of the theory as well as to the off-shell BRST operator so that the
quantization can be done in the standard way, i.e. as in Yang-Mills type
theories. Let us note that we first apply our prescription to theories
described by a gauge algebra with vanishing higher-order gauge functions,
i.e. theories of type $(2,2)$ which contain all the subtleties required to
the insight of the procedure. Then a direct generalization is given for any
open gauge theory of type $(p,q)$, with, however, particular technical
remarks that stand out in the general case. In the last chapter we study the
particular problem of the minimal set of auxiliary fields for any given open
gauge theory. Then we end up with a quick formulation of the procedure for
simple supergravity and reproduce the standard results.

To quantize gauge systems, the exposed prescription should be compared to
the BV approach. The latter is not the unique way to quantize closed and
irreducible gauge theories but became impossible to circumvent for open
and/or reducible theories for the reason that no systematic procedure for
the introduction of auxiliary fields was to date available. At first sight,
the comparison clearly stops at the on-shell level for the reason that in
the BV procedure, the nilpotency of the BRST operator is guaranteed only on
shell after the elimination of the antifields. It is worth noting that at
this level both of the two procedures leads to the same higher-order ghost
coupling terms in the on-shell full quantum action. However, to step forward
and really quantize the theory, one may remark that a systematic procedure
for the introduction of auxiliary fields closes the classical algebra and
makes the quantum theory much simpler, since in this case the transformation
laws are linear\footnote{%
One can check from (73-74) that upon replacing the ghost fields by gauge
parameters one can easily see that the obtained transformations are linear.}
and lead to an off-shell BRST operator together with a complete off-shell
invariant action containing all the gauge fixing conditions. One can then
easily derive the so-called Ward identities which are necessary in many
aspects of the quantized theory, for instance gauge independence of the
partition function as well as perturbational proofs of unitarity and
renormalisability are heavily based on these identities \cite{14}. Then, to
determine the quantum theory completely one has to add an extra symmetry,
i.e. the so-called shift symmetry upon introducing the set of collective
fields in order to obtain the quantum equations of motion, i.e. the
Schwinger-Dyson equations \cite{15} (see also Ref.\cite{16}) as Ward
identities of the complete theory and end up with a physical quantum theory,
in the sense that all the physical degrees of freedom are fixed, together
with an off-shell structure of the symmetries. This can not be realized in
the BV quantization scheme. In this approach, in order to obtain a theory
with all the fixed degrees of freedom, one has to require the elimination of
the antifields for the benefit of the gauge fixing functions trough the
gauge fixing fermion and this leads inevitably to an on-shell structure of
the symmetries. But if we want to quantize the theory effectively and derive
the Ward identities one has to reintroduce the antifields and take advantage
of the off-shell structure provided by this reintroduction. One can then
clearly see that in the BV formalism, a physical quantum theory can not be
obtained together with an off-shell structure contrary to what can be done
via the introduction of auxiliary fields that realizes the off-shell
nilpotency and allows in the same time the introduction of all the gauge
fixing functions without introducing any new physical degrees of freedom in
the sense that they are non-propagating fields. Let us also remark that
besides the fact that auxiliary fields simplify greatly the quantization of
open gauge theories, they are of particular interest in many specific cases.
For example, one can cite globally supersymmetric models such as the
Wess-Zumino model, for which it is only with auxiliary fields that one can
obtain a tensor calculus \cite{3}. One can also mention the case of BF
theories which represent models of reducible theories but have, however, an
on-shell structure, and for which the introduction of auxiliary fields
realizes the metric independence of the BRST operator and allows to simplify
the proof of the metric independence of the partition function of such
theories \cite{4}.

However, one should mention that an interesting idea exists in order to
extend the BV method to investigate a possible realization of a complete
off-shell quantization procedure. In their approach (see Refs. [11-13]) the
authors are led to identify the auxiliary fields through the variation of
the gauge fixing fermion with respect to the gauge fields of the classical
theories. This method leads at first sight to three binding remarks. The
first one concerns the non vanishing ghost number of the auxiliary fields
obtained in this way. This clearly compromise the possibility to considering
these fields at the classical level and thus jeopardize the construction of
a classical extension of the theory. The second remark is related to the
particular constraints taken by the authors on the gauge functions of the
gauge algebra. These constraints imposed for internal consistency reduce
considerably the logical simplicity of the theory and potential
generalizations (see in particular Ref. \cite{11}). The last remark affects
the representation (and then the number) of the auxiliary fields. Indeed, in
their approach we see that these fields are inevitably in the same number
that of the gauge fields with non vanishing ghost number and opposite
statistic. This can rise the problem of the definition of the minimal set of
auxiliary fields. Nevertheless, in Ref. \cite{13} the authors bring a clever
way to bypass this difficulty for the specific case of simple supergravity,
but they take too much advantage of the particularities of the theory to
envisage a smooth generalization to general open gauge theories. As a quick
comparison, our prescription gives rise to auxiliary fields with vanishing
ghost-numbers and their representation is only related, upon the
field-redefinition (39), to the symmetries of the classical theory. This
permits us in Sec. IV to analyze the question of the minimal representation
in a general framework. The constraints used in this section are twice. The
first and more important one (44) is a very general condition related to the
nature of any set of auxiliary fields that impose to them that they must not
introduce any new degrees of freedom to the classical theory and this
condition finds its theoretical meaning in the very general explicit
function theorem \cite{9}. The second condition (53) is related to the gauge
fixing functions that are taken to have not any kind of invariance, which is
not a strong restriction in virtue of the freedom in fixing the gauge.

Finally, one should mention that in order to study all further possible
advantages of the auxiliary fields structure, it would be interesting to
reinvestigate the prescription presented in this paper in a more formal way%
\footnote{%
The essential motivation of the present paper was to show how the
introduction of auxiliary fields can be practically realized for irreducible
open gauge theories.} and also to find out how to make a generalization to
reducible gauge theories. Furthermore, to develop and consolidate our
approach, it would be also interesting to apply it for several specific
theories. In particular, we plan to use it to give a complete off-shell
formulation of the eleven-dimensional (11D) supergravity for which the
complete structure of the auxiliary fields is unknown. Let us note that 11D
supergravity recently became interesting because of its return in the
so-called M-theory (for a review see Ref.\cite{17}) and only a partial
off-shell formulation has been already proposed in Ref.\cite{18}.

\begin{acknowledgement}
M.T. acknowledges support from the Alexander von Humboldt foundation in the
framework of the Georg Forster program.
\end{acknowledgement}

\end{document}